\documentstyle[epsf]{l-aa} 

\begin{document}

\newcommand{\mic}{\,{\rm \mu m} } 

\thesaurus{Sect. 02: cosmology ,
           12.04.2,     % Cosmology: diffuse radiation
           13.09.3,     % Infrared:ISM:continuum 
           13.09.2}     % IR:general

\title{Detection of the extra-Galactic background fluctuations at 170~$\mu$m\thanks{Based 
on observations with ISO, an ESA project with instruments
funded by ESA Member States (especially the PI countries: France, Germany,
the Netherlands and the United Kingdom) and with the participation of ISAS
and NASA}}

\author{G. Lagache \and
	J.L. Puget}

\institute{Institut d'Astrophysique Spatiale, B\^at.  121, 
Universit\'e Paris XI, F-91405 Orsay Cedex, France}
 
\offprints{lagache@ias.fr}

\date{Received, 18 march 1999; Accepted, 10 october 1999}  
          
\maketitle

\begin{abstract}
We have used the Marano1 field observations with ISOPHOT at 170 $\mu$m
to search for the Cosmic Far-InfraRed Background
fluctuations. This field is part of the FIRBACK 
project (Puget et al. 1999). For the
first time, fluctuations due to unresolved extra-Galactic sources
are isolated. The emission of such sources clearly dominates,
at arcminute scales, the background fluctuations
in the lowest Galactic emission regions.\\

The study presented here is based on a power spectrum analysis
which allow to statistically separate each of the background components. 
With this analysis, we clearly show that we detect the CFIRB
fluctuations.


\end{abstract}

\section{Introduction}
The major component of the extra-Galactic background is formed by the 
integrated light of all distant galaxies that are not resolved in the beam
of the instrument. 
The detection of the far-infrared part of this background
(the CFIRB, Cosmic Far-InfraRed Background) with COBE
(Puget et al. 1996; Fixsen et 
al. 1998; Hauser et al. 1998; Lagache et al. 1999)
has opened new perspectives in our understanding
of galaxy evolution. The energy contained in the CFIRB, compared
to the one in the UV/optical/near-IR domain (Lagache 1998; Gispert
et al. in preparation)
shows that a new population of galaxies has to be considered
in the general framework of galaxy evolution: the far-infrared
galaxies. The closest and/or most luminous ones have been discovered
by IRAS (see for a review Sanders \& Mirabel 1996). Nowadays, the
IR galaxies are hunted down
by several deep cosmological surveys such
as the ones with SCUBA at 850 $\mu$m 
(Hughes et al. 1998; Smail et al. 1997, 1998; Lilly et al. 1999;
Barger et al. 1999), ISOPHOT at 170~$\mu$m 
(Kawara et al. 1998; Puget et al. 1999)
and ISOCAM at 15 $\mu$m (Oliver et al. 1997; Aussel et al. 1999; D\'esert et al. 1999;
Elbaz et al. in preparation).\\

FIRBACK is a deep cosmological survey dedicated to the investigation of
the nature of the galaxies that contribute to the CFIRB
at 170 $\mu$m (Dole et al. 1999)
with the ISOPHOT instrument (Lemke et al. 1996).
It covers about 4 square degrees. The first FIRBACK field
to be observed and analysed is the Marano1 region centered on 
RA(J2000)=3h13m9.6s and DEC(J2000)=-55d03m43.9s. 
This field covers 0.25 square degree and 
contains 24 extra-Galactic sources with fluxes ranging from
100 to 850~mJy (Puget et al. 1999). It lies
in a very low cirrus emission region with HI column density equal
on average to 10$^{20}$~H~cm$^{-2}$. This low cirrus
contamination allows to search for brightness fluctuations 
due to unresolved sources. \\
In this paper, 
we use the FIRBACK Marano1 field to isolate the background fluctuations
with a high signal to noise ratio. We then discuss the separation of the background
fluctuations between the Galactic and extra-Galactic component.\\
Throughout this paper, ``background'' refers to the extra-Galactic
and Galactic backgrounds, and ``CFIRB'' to the extra-Galactic background
at far-IR wavelengths.

\section{Sources of the CFIRB}

The CFIRB is made of sources with number counts as a function of flux which can
be represented, for the present discussion, by a simple power law:
\begin{equation}
\label{eq_count}
N(>S)= N_0 \left( \frac{S}{S_0} \right)^{- \alpha}
\end{equation}
For an Euclidian Universe, $\alpha$=1.5.
Obviously, these number counts need to flatten at low fluxes
to insure a finite value of the background. Thus, we assume in the
present discussion that $\alpha$=0 for $S<S^{\ast}$.\\

The intensity of the CFIRB, induced by all sources with flux
up to S$_{max}$, is given by:
\begin{equation}
\label{eq_cfirb_int}
I_{CFIRB}= \int_{0}^{S_{max}} \frac{dN}{dS} S dS
\end{equation}
For the simple Euclidian case ($\alpha$=1.5), the CFIRB integral 
is dominated by sources near S$^{\ast}$.\\

Fluctuations from sources
below the detection limit S$_0$ (which corresponds to the flux of sources 
either at the confusion or at the sensitivity limit) are given by:
\begin{equation}
\sigma^{2}= \int_{0}^{S_0} S^2 \frac{dN}{dS} dS \quad \mathrm{Jy^{2}/sr}
\end{equation}
Computing $\frac{dN}{dS}$ using Eq. (\ref{eq_count}), this gives:
\begin{equation}
\label{eq_sigmaa}
\sigma^{2}= \frac{\alpha}{2-\alpha}N_0 S_0^2 
\left[ 1- \left(\frac{S^{\ast}}{S_0} \right) ^{2-\alpha} \right] \quad \mathrm{Jy^2/sr}
\end{equation}
For the simple Euclidian case , the CFIRB fluctuations are dominated by sources 
which are just below the detection limit S$_0$. \\

Nevertheless, it is well known that strong cosmological
evolution, associated with a strong negative K-correction, could lead to a very steep
number count distribution (see for example Guiderdoni et al. 1998; Franceschini et al. 
1998). For $\alpha>$2, 
the CFIRB integral is still dominated by sources near S$^{\ast}$ but
its fluctuations are now also dominated by sources close to 
S$^{\ast}$.
In such a case, observations of the fluctuations will constrain
sources that dominate the CFIRB. 
Futhermore, investigations of problems like the fraction of AGN contributing
to the CFIRB, which has been identified as a major question of the coming years,
cannot be conducted only with individually  detected sources.
These observed sources correspond to the bright
part of the luminosity function and only represent a few percent of the
CFIRB. Moreover, the fraction of AGNs could very well be 
increasing with luminosity
(Sanders \& Mirabel 1996). Thus
the only way to tackle this problem will be to study 
the correlation between the extra-Galactic background fluctuations 
in the X-rays and Far-IR domains.\\

In the Far-IR, present observations show a very steep slope
$\alpha$=2.2 (Dole et al. 1999). Sources detected
above S$_{0}$ contribute only to less than 10$\%$ of the CFIRB
(with S$_{0}$=120 mJy). Thus, it is essential to study the 
extra-Galactic background fluctuations which are likely to be dominated
by sources with a flux comparable to those dominating the
CFIRB intensity.

\begin{figure}  
\epsfxsize=9.cm
\epsfysize=7.cm
\epsfbox{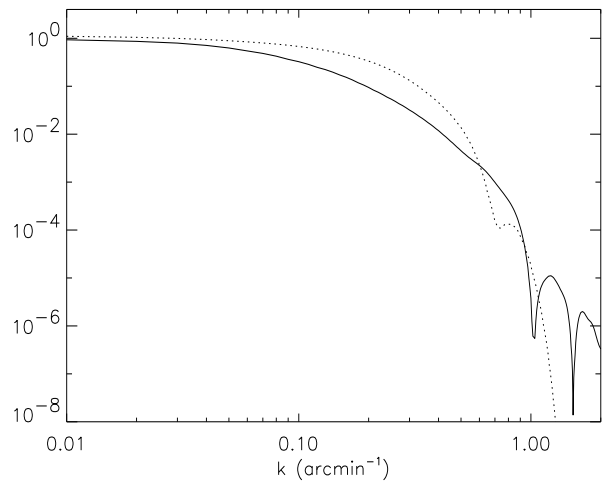}
\caption{\label{footprint} Continuous line: ISOPHOT Saturn footprint
power spectrum (Ws$_k$). Dotted line: ISOPHOT model
footprint power spectrum (Wm$_k$). The difference in
solid angle between the two footprints is equal to 25$\%$.}
\end{figure}

\begin{figure*}
\begin{minipage}{8.2cm}
\epsfxsize=9.cm
\epsfysize=9.cm
\hspace{0.5cm}
\vspace{0.4cm}
\epsfbox{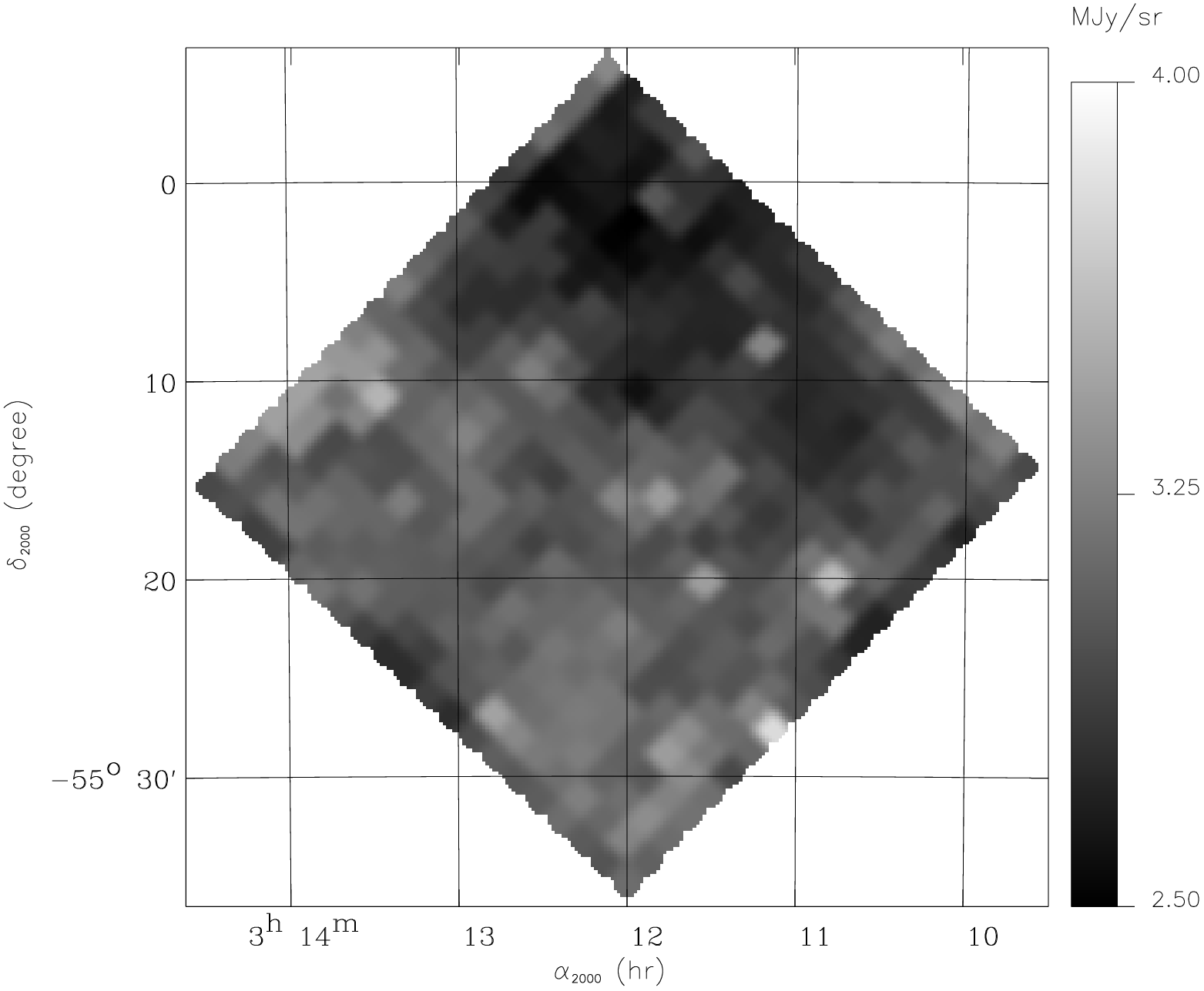}
\end{minipage}
\begin{minipage}{8.5cm}
\epsfxsize=9.cm
\epsfysize=9.cm
\hspace{2.cm}
\vspace{0.4cm}
\epsfbox{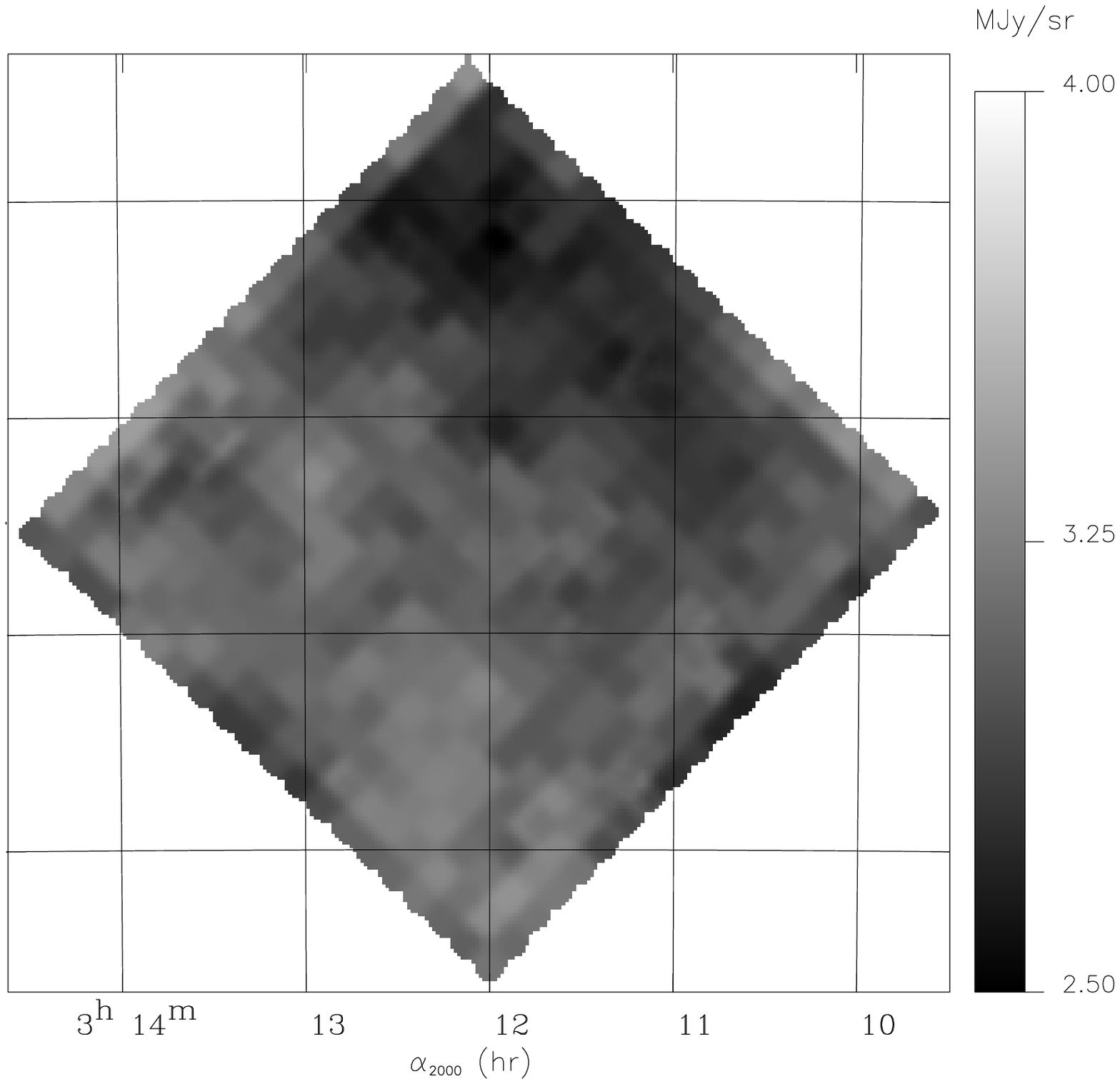}
\end{minipage}\\
\vspace{-1.cm}
\caption{\label{var_hi} a. The Marano1 region observed with ISOPHOT at 170~$\mu$m.
b. same as (a) with detected sources removed.}
\end{figure*}

\section{Summary on the data reduction and calibration}
The data reduction and calibration for this field
have been detailed in Lagache (1998), 
Puget et al. (1999) and Puget \& Lagache (1999).
We are just going to summarise here the different steps.\\

We use PIA, the ISOPHOT interactive analysis software 
version 6.4 (Gabriel et
al. 1997), to correct for instrumental effects, glitches induced
by cosmic particles and to provide an initial calibration. First we apply
the non-linearity correction. Deglitching is performed for each individual
ramp and then the mean signal per position is derived by averaging
the linear fit on each ramp (see Gabriel et al. 1997 for details).
The signal shows a long term drift
which is corrected by using the two internal calibration
(FCS) measurements bracketing each individual observation. 
Inside each raster, the long term drift represents less than 10$\%$ 
of the signal. The calibration is performed by deriving the mean value of 
the two FCS measurements. The contribution of the long term drift 
is thus lower than 5$\%$. Flat fielding is achieved using the high 
redundancy in the measurements.
Glitches inducing long term drifts are also corrected.
The flux is
finally projected on a 10"x10" coordinate grid using our
own projection procedure.\\

The final calibration is achieved using the footprint measured on Saturn.
It is presented in Fig. 1 of Puget et al. (1999). Brightnesses derived
using this footprint are in very good agreement with DIRBE.
The footprint measured on Saturn has been compared to the one
derived from the diffraction model, including the tripode
(Klaas et al. in preparation). The model takes into account the bandpass filter
and is computed for a $\nu$I$_{\nu}$=constant spectrum.
Thus, comparison between the model and Saturn
footprints allows to test the wavelength dependence
of the footprint. The solid angle supported by the two footprint differs
only by 25$\%$.
Comparison of the two footprints power spectra, Ws$_{\mbox{k}}$
(Saturn) and Wm$_{\mbox{k}}$ (model), is shown on Fig. \ref{footprint}. 
The Saturn FWHM footprint is quite larger than the model one.
The difference between these power spectra will be considered in
Sect. \ref{pest}.\\

The Marano1 observations consist of four 19x19 rasters. 
Each raster was performed in the spacecraft (Y,Z) coordinate system which
is parallel to the edges of the detector array. The field area covered by each
raster is about 30'x30'. The exposure time is 16s per pixel.
Rasters were performed with one pixel overlap in
each Y and Z direction. Therefore, for the maximum redundancy region, the
effective exposure time is 256 sec per sky position (16 sec per pixel,
4 rasters and redundancy of 4).
Displacements between the four raster centers correspond to 2 pixels. This mode of
observation, which does not provide proper sampling of the point-spread-function,
was chosen deliberately because it allows a very clean determination of the instrumental noise.\\

The final map is made of 16 coadded independent maps. 
The extraction of extra-Galactic sources and determination of their
fluxes are made on this final coadded map. 
To compute the instrumental
noise, we have measured the flux of the detected sources on the 16
independent maps. We then compute for each source the standard deviation
with respect to the flux measured on the coadded map. The average
of all the standard deviations gives the instrumental noise.
We obtain 1.74~mJy rms in a 89.4''$\times$89.4'' pixel. 
The signal in the map is around 3~MJy/sr. This leads to a very high signal to
noise ratio in the map of about 300.\\

Using the four maps of the Marano1 field presented in Fig. 2 of Puget et
al. (1999), we can make an estimate of
the instrumental noise power spectrum, making the
difference between the independent maps. We know that
this power spectrum contains low spatial
frequency noise, due to small distorsions in the
flat field and slow response changes during the observations.
With such map differences, we obtain instrumental noise power spectra
for pairs of maps which 
are all more than 10 times smaller than the measured 
total flux power spectum (which is presented in Fig. 3).
The instrumental noise power spectrum is rather flat in the range k=[0.1, 0.5] arcmin$^{-1}$ 
(about 30 Jy$^2$/sr at k$\sim$0.2 arcmin$^{-1}$ and 10 Jy$^2$/sr at k$\sim$0.4 arcmin$^{-1}$) and 
is in very good agreement with the one deduced from
the noise (assumed to be white) computed on the extra-Galactic sources 
(1.74~mJy rms in a 89.4''$\times$89.4'' pixel). 
The measured instrumental noise power spectrum is subtracted to the
measured power spectrum before the analysis.

\section{The extra-Galactic sources}

In the Marano1 field, 24 extra-Galactic sources are detected with fluxes ranging from
100 to 850 mJy (Puget et al. 1999). Sources have been extracted up to the confusion 
limit (3$\sigma$=67.2 mJy) and source fluxes been computed using the
Saturn footprint measurement for which a significant fraction of the integral
is in the wing (and which is in good agreement with the footprint
model). The source subtraction from the Marano1 map
is very straightforward. We subtract
at each source position the footprint, sampled at 10 arcsec and normalised
to unit integral, multiplied by the source flux.
The map obtained after the source subtraction is given Fig.~2 together
with the original map. The standard deviation in the source subtracted map is 
around 0.14 MJy/sr, which is very close to the one of the original map which is about
0.15 MJy/sr. For the mean brightness, the difference between the two maps is
less than 1$\%$. These numbers clearly show that the background
fluctuations are not dominated by strong sources detected above the confusion
limit. As for the original map, the source subtracted map has
a very high signal to noise ratio.\\

\begin{figure*}  
\epsfbox{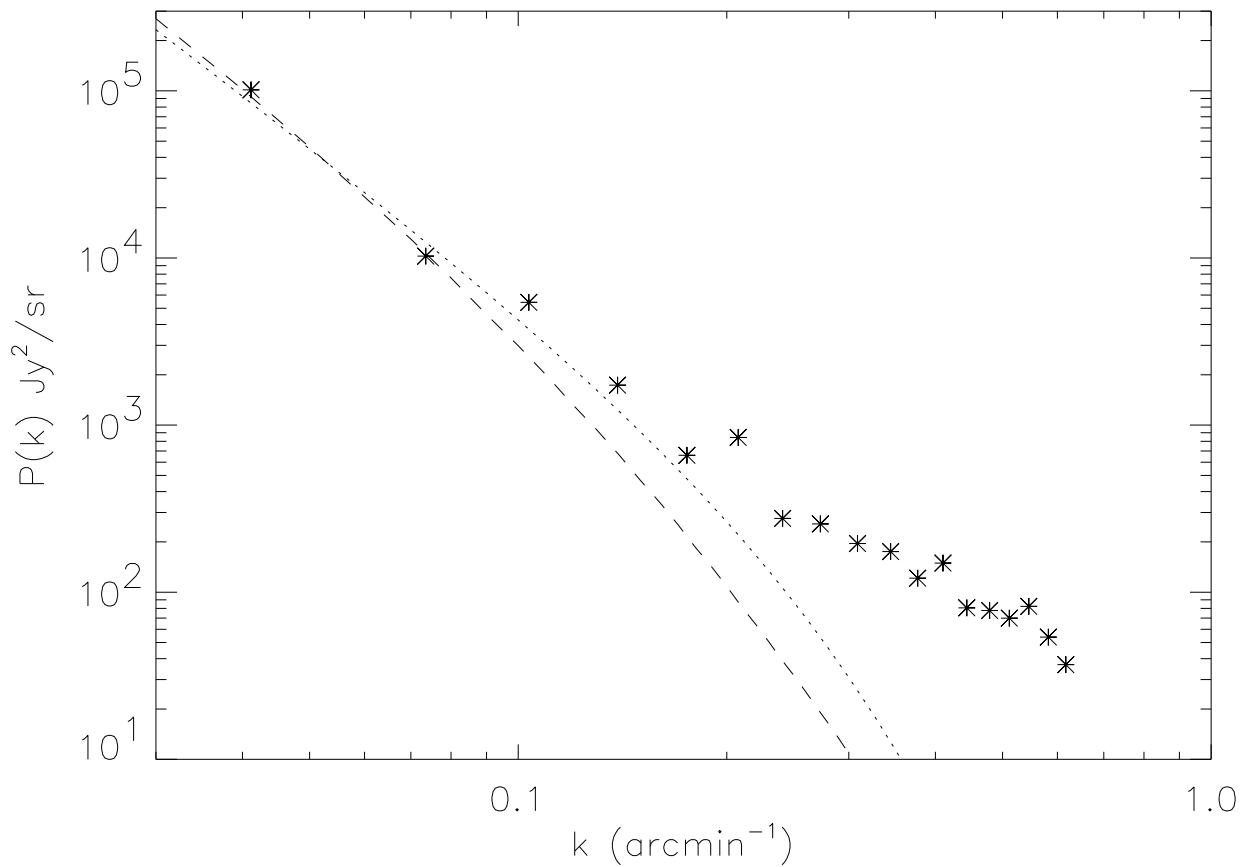}
\caption{Power spectrum of the source subtracted map ($\ast$).
The dashed and dotted lines correspond to the
cirrus confusion noise multiplied by
the footprint power spectrum (Ws$_{\mbox{k}}$ and Wm$_{\mbox{k}}$ respectively)}
\end{figure*}

Background fluctuations (Fig. 2b) are made of two components: the Galactic cirrus
one and the extra-Galactic one
(the instrumental noise is negligible and is not correlated).
We are going now to discuss the relative contribution of these two components.

\section{Extra-Galactic and Galactic fluctuation separation}

Our separation of the extra-Galactic and Galactic fluctuations
is based on a power spectrum decomposition. This method
allows us to discriminate the two components using the statistical
properties of their spatial behaviour. \\

We know from previous works that the cirrus far-infrared 
emission power spectrum
has a steep slope in $k^{-3}$ (Gautier et al. 1992; Kogut et al. 1996; 
Herbstmeier et al. 1998; Wright 1998). These observations
cover the relevant spatial frequency range. Futhermore,
Abergel et al. (1999) have shown that this spatial structure,
measured from extinction, extend to arcsecond scales.
Despite a large observationnal effort, no characteristic scale
has been identified in the diffuse interstellar medium (Falgarone 1998).\\
For the extra-Galactic fluctuations, little is known.
The power spectrum of galaxy clustering
has been mainly studied from optical surveys.
The main result is that the power spectrum
can be represented by a simple power law with an index
equal to -1.3 at spatial frequency larger than $\sim$0.1 degree$^{-1}$
(Groth \& Peebles 1977; Maddox et al. 1990; Peacock 1991;
Kashlinsky 1992).
For small spatial frequency, if we assume the standard
inflationary model ($\Omega$=1 and the Harrison-Zeldovich power
spectrum on large scales), the galaxy power
spectrum is proportional to k.
At Far-IR wavelengths, nothing is known about the galaxy
clustering. A spatial and redshift distribution similar
to the optical surveys (as the ``Automatic Plate Measuring'' survey) will give
a negative slope around -1.3 whereas a Poissonian distribution
will give a flat power spectrum.\\

In summary, we know that the cirrus emission power spectrum
is steep and does not present any characteristic scale. The extra-Galactic
component is unknown but certainly much flatter. We thus conclude
that the steep spectrum observed in our data at k$<$0.15 arcmin$^{-1}$
(Fig. 3)
can only be due to the cirrus emission. The break in the power spectrum
at k$\sim$ 0.2 arcmin$^{-1}$ is very unlikely to be due to the
cirrus emission itself which is known not to exhibit any
prefered scale.  
Thus, the power spectrum is a powerful tool for a first order separation
between the Galactic and extra-Galactic background component.
We now give and discuss the quantative separation of the
two components.

\subsection{Normalisation of the cirrus emission fluctuations}
The cirrus emission has a filamentary structure which, as discussed above,
exhibits a spatial power spectrum which has been first
determined at 100~$\mu$m from IRAS data by Gautier et al. (1992):
\begin{equation}
\label{PS_Gautier}
P_{\mbox{cc}, 100}(k)= 1.4 \times 10^{-12} {B_0}^{3} \left( \frac{k}{k_0} \right)^{-3} \mathrm{Jy^2 /sr}
\end{equation}
with k$_0$ = 10$^{-2}$ arcmin$^{-1}$, B$_0$ the cirrus brightness at
100~$\mu$m in Jy/sr and k the spatial frequency in arcmin$^{-1}$.
The main problem in this determination comes from the normalisation
which has been established in high
cirrus emission regions. This normalisation does
not necessarily apply to our very low cirrus emission
field. For this reason, the normalisation is directly determined
using the measured power spectrum (Fig. 3). Using
the low frequency data points and assuming a
$k^{-3}$ dependance, we find a power spectum of:
\begin{equation}
P_{\mbox{cc}, 170}(k)= 7.9^{+2.1}_{-2.7} \quad k^{-3} \quad \mathrm{Jy^2/sr}
\end{equation}
with k in arcmin$^{-1}$.
The main uncertainty comes from the footprint
used for deriving $P_{\mbox{cc}, 170}(k)$.
The normalisation of the power spectrum derived from
the low frequency data points agrees reasonably well
with Wright (1998) DIRBE results at high Galactic latitude.
It is also in good agreement with the Gautier et al. (1992)
normalisation scaled to 170 $\mu$m (Appendix A).

\begin{figure}
\epsfxsize=9.cm
\epsfysize=7.cm
\epsfbox{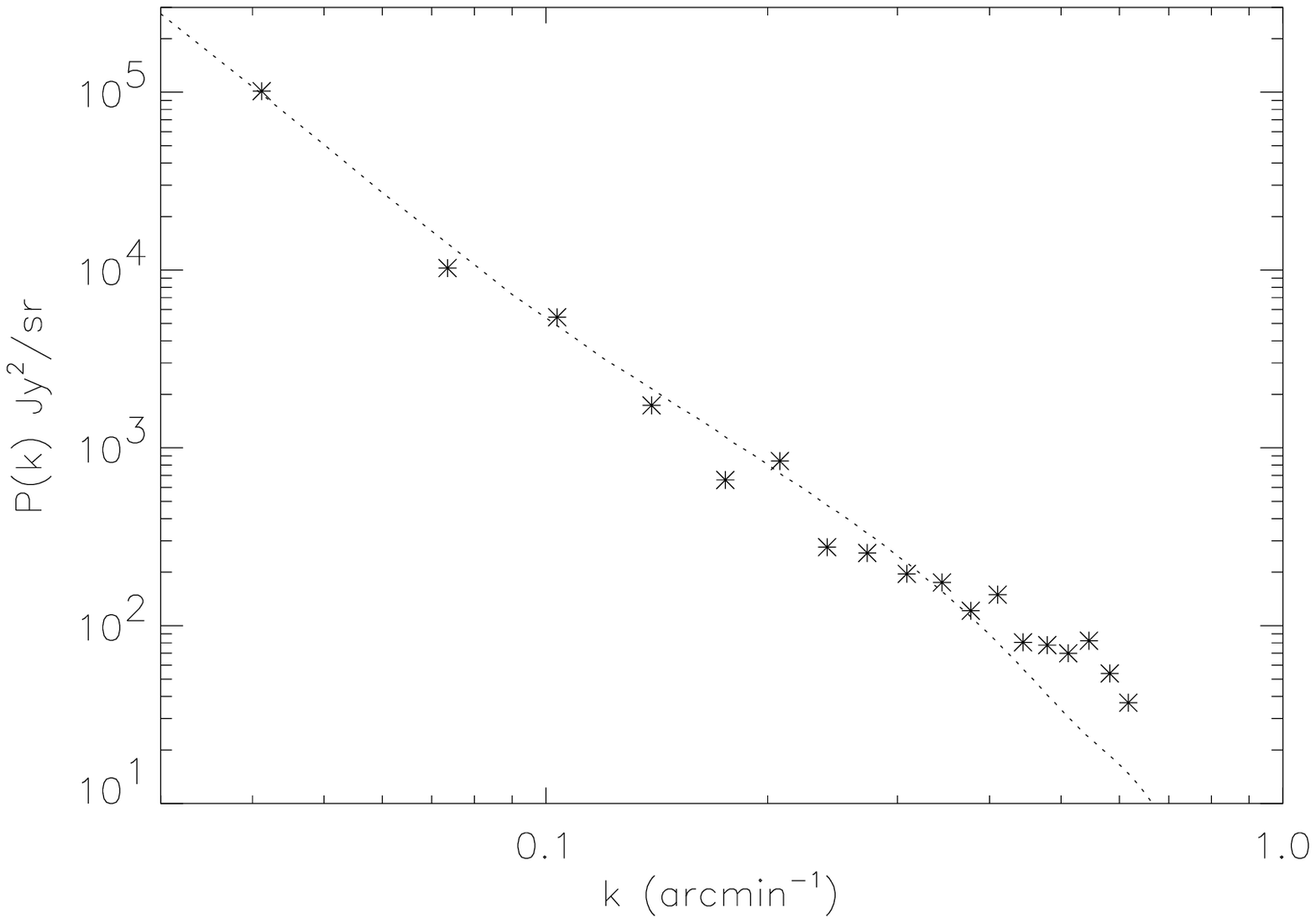}
\caption{$\ast$: power spectrum measured on the map.
Dotted line: total power
spectra obtained by summing up the cirrus confusion noise (adjusted
to our large scale points) and a flat CFIRB fluctuation power spectrum
P$_{\mbox{cs}}$~=~7400 Jy$^2$/sr (we have used here 
the Saturn footprint power spectrum).}
\end{figure}

\subsection{\label{pest} Measured power spectrum}
We can now compare the power spectrum measured on the source subtracted map
(Fig. 1b)
with our estimated one for the non cosmological components
defined as:
\begin{equation}
\label{Pest_eq}
P_{\mbox{est}}= P_{\mbox{cc}, 170} \times W_{\mbox{k}} 
\end{equation}
where  W$_{\mbox{k}}$ is the ISOPHOT 170 $\mu$m beam power spectrum,
and P$_{\mbox{cc}, 170}$ the cirrus confusion noise.
Fig. 3 shows the measured power spectrum 
together with P$_{\mbox{est}}$ computed for
W$_{\mbox{k}}$=Ws$_{\mbox{k}}$ and W$_{\mbox{k}}$=Wm$_{\mbox{k}}$ 
(as defined in Fig. \ref{footprint}). 
We clearly see an excess between
k=0.25 and 0.6 arcmin$^{-1}$ which is more than a factor 10
at k=0.4 arcmin$^{-1}$. Any reasonable power law spectrum
for the cirrus component multiplied by the footprint
leads, as can be easily seen in Fig. 3, to a very steep
spectrum at spatial frequency k$>$0.2 arcmin$^{-1}$.
This is very different from the observed spectrum.
The excess is quite independent of
the footprint used. Moreover, it is more than
10 times larger than the measured instrumental noise power spectrum.
Therefore, as no other major source of fluctuations
is expected at this wavelength, the large excess observed between
k=0.25 and 0.6 arcmin$^{-1}$ is interpreted as due
to unresolved extra-Galactic sources.
The anisotropies of the background are thus dominated, at angular scales
close to the angular resolution, by the structure of the CFIRB.
This is the first detection of the fluctuations of the CFIRB.\\

CFIRB fluctuations are clearly detected between
k=0.25 and 0.6 arcmin$^{-1}$. This interval has
not enough dynamic range to constrain the slope of the
CFIRB fluctuation power spectrum:
our present study does not allow to clearly constrain the clustering
of galaxies. However, the power spectrum mean level can be determined.\\
Power spectra with simple power law with index smaller than -1 are
not compatible with our measured power spectrum. 
For indexes in the range [-1,0], 
the different power laws give equivalent results.
For example, on Fig. 4 is shown the total power
spectrum obtained by summing up the cirrus confusion noise 
and a constant CFIRB fluctuation power spectrum,
P$_{cs}$~=~7400 Jy$^2$/sr.
This white noise power spectrum level is in very good
agreement with the one predicted by Guiderdoni et al. (1997). 
This gives CFIRB rms fluctuations around 0.07 MJy/sr (for
a range of spatial frequency up to 5 arcmin$^{-1}$).
These fluctuations are at the $\sim$9 percent level,
which is very close to the predictions of Haiman
\& Knox (1999).

\section{Conclusion}
We have shown in the Marano1 region (30'$\times$30') that the 
extra-Galactic background
fluctuations are well above the instrumental noise
and the cirrus confusion noise. The observed power spectrum shows a flattening
at high spatial frequencies
which is due to extra-Galactic unresolved sources.
A preliminary study on the FIRBACK N1 field (2 square degrees)
show exactly the same behaviour. 
The statistical analysis of the 
background fluctuations using the full FIRBACK fields,
covering different high Galactic latitude
regions, will allow to constrain models of galaxy evolution
and will be a very useful tool for cosmological studies
until large passively cooled telescopes in space become
available which will allow number counts down to a few
mJy.
The next step consists of removing the cirrus contribution using gas tracers
(the H$_{\alpha}$ and the 21cm emission lines)
to isolate the extra-Galactic fluctuation
brightness and thus constrain the clustering of galaxies.\\

%=========================================================================
Acknowledgements:\\
We would like to thanks A. Abergel, R. Gispert, F.R. Bouchet for many
useful discussions and the ISOPHOT team for many interactions on the data reduction.
Thanks to H. Dole for the ISOPHOT footprint model.
The work presented here profited from very useful comments from the referee.\\

{\bf Appendix A: Comparaison of the measured cirrus power spectrum with the
Gautier et al. (1992) normalisation:}\\

The power spectrum of the cirrus emission derived by Gautier et al. (1992)
from IRAS data at 100~$\mu$m follows:
\begin{equation}
\label{PS_Gautier}
P_{\mbox{cc}, 100}(k)= 1.4 \times 10^{-12} {B_0}^{3} \left( \frac{k}{k_0} \right)^{-3} \mathrm{Jy^2 /sr}
\end{equation}
with k$_0$ = 10$^{-2}$ arcmin$^{-1}$, B$_0$ the cirrus brightness at
100~$\mu$m in Jy/sr and k the spatial frequency in arcmin$^{-1}$.\\

The far-infrared Galactic sky at high latitudes has essentially three
components: the zodiacal background, the emission from
Galactic cirrus and the emission from infrared galaxies
at all redshifts.
The average total flux in our map is 3.16 MJy/sr.
The zodiacal emission has been derived at the time
of our observation using the model
of Reach et al. 1995 (calibrated on DIRBE data)
and is equal to 0.67$\pm$0.07 MJy/sr.
Following Lagache et al. (1999), the CFIRB at 170 $\mu$m is equal to
0.77$\pm$0.25 MJy/sr. Thus, we obtain for the cirrus emission at
170~$\mu$m, I(170)=1.72$\pm$0.26 MJy/sr.
Using the color ratio for the HI diffuse
emission I(170)/I(100)=1.96 (Lagache et al. 1999), 
we derive B$_{0}$= 0.88$\pm$0.13 MJy/sr= 8.8$\pm$1.3 10$^5$ Jy/sr and
the normalisation of the power spectrum at 170 $\mu$m becomes:
\begin{equation}
1.4 \times 10^{-12} \left[\frac{I(170)}{I(100)} \right] ^2 B_{0}^{3}= 3.63^{+1.9}_{-1.4} \times 10^6 \quad \mathrm{Jy^2/sr}
\end{equation}
Using Eq. (\ref{PS_Gautier}), this gives:
\begin{equation}
P_{\mbox{cc}, 170}(k)= 3.63 ^{+1.9}_{-1.4} \times k^{-3}  \quad \mathrm{Jy^2/sr}
\end{equation}
with k in arcmin$^{-1}$.\\

In fact Gautier et al. results fit to within a factor of 2
the Eq. (\ref{PS_Gautier}).
Taking into account this uncertainty, the
cirrus power spectrum becomes:
\begin{equation}
\label{eq_8}
P_{\mbox{cc}, 170}(k)= 3.63 ^{+7.46}_{-2.51} \times k^{-3}  \quad \mathrm{Jy^2/sr}
\end{equation}
We do not expect in such a low column density field 
(N$_{HI} \sim$ 10$^{20}$ H cm$^{-2}$) any
contribution of dust associated with molecular gas (Lagache
et al. 1998). Therefore, we can argue that the normalisation
of the cirrus confusion noise at 170~$\mu$m, for
our field, is between 1.1 and 11.1 Jy$^2$/sr.\\

Our normalisation computed from the low frequency data
points is around 7.9 Jy$^2$/sr which is in the
range of Gautier et al. results. This
value of 7.9 Jy$^2$/sr agree also reasonably well
with Wright (1998) measures using DIRBE data.


\begin{thebibliography}{}
\bibitem{abergel99} Abergel A., Andr\'e P., Bacmann A., et al., 1999
in The Universe as seen by ISO, ESA-SP 427
\bibitem{Aussel} Aussel H., Cesarsky C.J., Elbaz D., Starck J.L., 1999, 
A\&A in press
\bibitem[1999]{Barger}Barger A.J., Cowie L.L., \& Sanders D.B., 1999, ApJ 518, L5
\bibitem{desert98} D\'esert F.X., Puget J.L., Clements D., et al., 
1999, A\&A in press
\bibitem{dole99} Dole H., Lagache G., Puget J.L., et al., 1999,
in The Universe as seen by ISO, ESA-SP 427
\bibitem{Falga} Falgarone E., in Starbursts: triggers, nature and evolution,
Les Houches School, 1998, Ed. B. Guiderdoni, A. Kembhavi
\bibitem{Fixsen98} Fixsen D.J., Dwek E., Mather J.C., et al., 1998, ApJ 508, 123
\bibitem{Franceschini} Franceschini A., Andreani P., Danese L., 1998, MNRAS 296, 709
\bibitem{Gabriel} Gabriel C., Acosta-Pulido J., Henrichsen I., et al., 1997, ADAAS VI, A.S.P. Conference
series, vol. 125, G. Hunt, H.E. Payne eds, p 108
\bibitem{gautier} Gautier T.N.III, Boulanger F., P\'erault M., 
Puget J.L., 1992, AJ 103, 1313
\bibitem{Guider} Guiderdoni B., Bouchet B., Puget J.L., et al., 1997, Nature 390, 257
\bibitem{guider98} Guiderdoni B., Hivon E., Bouchet F.,
Maffei B., 1998, MNRAS 295, 877
\bibitem{Groth} Groth E.J., Peebles P.J.E., 1977, ApJ 217, 385
\bibitem{Haiman} Haiman Z., Knox L., 1999, ApJ in press
\bibitem{Hauser98} Hauser M.G., Arendt R.G., Kelsall T., et al., 1998, ApJ 508, 25
\bibitem{herbs} Herbstmeier U., Abraham P., Lemke D., et al., 1998, A\&A 332, 739
\bibitem{hughes98} Hughes D.H., Serjeant S., Dunlop J., et al., 1998, Nature 394, 241
\bibitem{Kashli} Kashlinsky A., 1992, ApJ 399, L1
\bibitem{Kawara} Kawara K., Sato Y., Matsuhara H., et al.,
1998, A\&A 336, L9
\bibitem{Kogut} Kogut A., Banday A.J., Bennett C.L., et al., 1996, ApJ 460, 1
\bibitem{laga} Lagache G., 1998, PhD thesis, University of Paris XI
\bibitem{lagac} Lagache G., Abergel A., Boulanger F., et al., 1998, A\&A 333, 709
\bibitem{lag} Lagache G., Abergel A., Boulanger F., et al., 
1999, A\&A 344, 322
\bibitem{Lilly99} Lilly S.J., Eales S.A., Gear W.K.S., et al., 1999, ApJ 518, 641
\bibitem{Lem} Lemke D., Klaas U., Abolins J., et al. 1996, A\&A 315, L64
\bibitem{Maddox} Maddox S.J., Efstathiou G., Sutherland W.J., Loveday J., 1990, MNRAS 242, 43
\bibitem{oliver} Oliver S.J., Goldschmidt P., Franceschini A., et al., 1997, MNRAS 289, 471
\bibitem{Pea} Peacock J.A., 1991, MNRAS 253, 1
\bibitem{Puget99b} Puget J.L., Lagache G., 1999, in The Universe
as seen by ISO, ESA-SP 427 
\bibitem{Puget96} Puget J.L., Abergel A., Bernard J.P., et al., 1996, A\&A 308, L5 
\bibitem{Puget99a} Puget J.L., Lagache G., Clements D.L., et al., 
1999, A\&A 354, 29
\bibitem{Reach95} Reach W.T., Franz B.A., Kelsall T., et al., 
1995, in Unveiling the Cosmic
Infrared Background, ed. E. Dwek, AIP Conf. Proc. 
\bibitem{Sanders} Sanders D.B., Mirabel I.F., 1996, ARA\&A 34, 749
\bibitem{Smail97} Smail I., Ivison R.J., Blain A. W., 1997, ApJ 490, L5
\bibitem{Smail98} Smail I., Ivison R.J., Blain A. W., Kneib J.P., 1998, ApJ 507, 21
\bibitem{Wright} Wright E.L., 1998, ApJ 496, 1
\end{thebibliography}
\end{document}